\documentclass[10pt, twocolumn]{article}
\usepackage[dvips]{graphicx}

\title{Quantum bit detector}

\author{M.\,P.\,Klembovsky$^{+,*}$,
M.\,L.\,Gorodetsky$^{+,*}$\/\thanks{e-mail:
gorm@hbar.phys.msu.ru}, Th.\,Becker$^*$, H.\,Walther$^*$ \\
\small $^+$ Faculty of Physics M.V.Lomonosov Moscow State
University, Leninskie Gory, 119992 Moscow, Russia \\
\small $^*$Max-Planck-Institut f\"{u}r Quantenoptik, 85748
Garching, Germany}

\begin{document}

\maketitle

\begin{abstract}
We propose and analyze an experimental scheme of quantum
nondemolition detection of monophotonic and vacuum states in a
superconductive toroidal cavity by means of Rydberg atoms.
\end{abstract}

\bigskip

One of the key directions of modern physics is the following
challenging problem of understanding the essence of the process of
quantum measurement. Of special interest in this respect are
experiments with individual quantum objects. Such applications as
quantum computing, quantum cryptography and quantum teleportation,
which have recently been attracting increasing attention
\cite{computing}, have their roots in this field. Quantum
measurements and particularly experiments on the interaction of
individual atoms and ions with the quantum field in a cavity are
usually associated with the optical domain. With the development
of the Rydberg atom technique, however, impressive results have
been obtained in the microwave region \cite{Meschede}. This
technique allows preparation and nondestructive (quantum
non-demolition) repetitive measurements of Fock states with a
small number of quanta in a high-Q superconductive cavity
\cite{Varcoe, Haroche}.

In 1994 V.B.Braginsky and F.Ya.Khalili \cite{Braginsky} proposed
an elegant scheme employing Rydberg atoms which allowed
nondestructive detection of vacuum and monophotonic states. The
idea of the experiment is to use a cavity with a geometry such
that the flying atom can interact twice with the field. A
composite resonator comprising two sandwiched coaxial leucosapphire disks
with whispering gallery modes was proposed initially with the
atoms flying inbetween near the surfaces of the disks along their diameter.
If the atom's velocity and the geometry are chosen such that the
interaction time takes one half of the Rabi cycle, then the atom
and the field may effectively exchange photons with a probability
close to 100\%. Dual interaction ensures that an atom leaves the
cavity unexcited in both cases, when the cavity is in vacuum and
one-photon state. The only difference is the state of the atom in
the central area of the cavity between the two interactions. It
was suggested initially \cite{Braginsky} that an inhomogeneous
d.c. field be applied in this region. This electric field detects
states nondestructively (state-dependent deflection). A simpler
scheme for realizing nondestructive state detection was proposed
later \cite{Wagner}, and we discuss here a practical scheme for a
quantum nondemolition (QND) quantum bit detector (QBD), based on
the initial idea \cite{Braginsky} with a toroidal superconductive
cavity instead of sapphire disks.

The scheme of the proposed experiment is given in Fig.1. A Rydberg
atom prepared in a particular state enters the toroidal
superconductive cavity. If initially the cavity is in the vacuum
state of the photon field $|0\rangle$, the state of the atom does
not change in the first interaction region and the central RF
field (resonant with auxiliary transition) lowers the atom state
to the auxiliary lowest level. If, however, the initial state of
the cavity is the one-photon state $|1\rangle$, the atom absorbs
the cavity photon. To provide 100\% absorption of the photon the
interaction time has to be equal to one half of the Rabi cycle
($\bar gL/v=\pi/2$, where $\bar g$ is the effective Rabi
frequency, $L$ is the interaction length and $v$ is the velocity
of the atom). When the atom is in the upper state, the central
field is not resonant and the state is unchanged. During the
second interaction the atom returns the photon back to the cavity.
In this way the information about the cavity quantum state is
recorded in the atomic state and the quantum state of the cavity
is not destroyed. The information recorded in the atomic state can
easily be read out by the state-selective detector based on field
ionization.

\begin{figure}
\includegraphics[width=0.47 \textwidth]{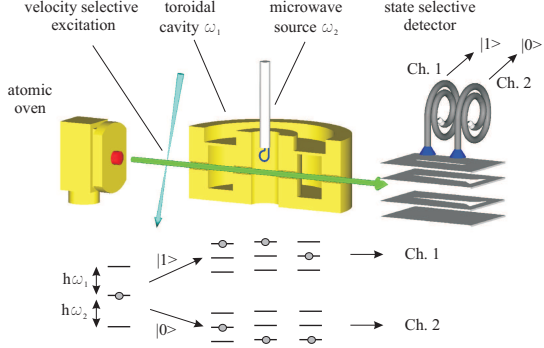}
\caption {\small Scheme of the quantum bit detector. The cavity is
shown in section.} \label{fig1}
\end{figure}

The transition between $61D_{5/2}$ and $63P_{3/2}$ levels of the
$^{85}Rb$ Rydberg states, the cavity mode (21.456 GHz) and
$61D_{5/2}\to 59F_{7/2}$ (21.122 GHz) for the probe are planned to
be used in the experiment.

The theory of the QBD based on the Jaynes-Cummings model and the
standard master equation approach can be created in the same way
as the theory of the micromaser field (see details in Refs.
\cite{Englert, Meystre, UFN}). The basic difference is that atoms
are prepared not in the excited but in the ground state and that
interaction between the cavity and the atoms is more complicated.

The complete initial state $P$ of the system atom in the ground
state plus the cavity field can be written in the form:
\begin{eqnarray}
P=\sum_{n,m}|g,n\rangle\rho_{nm}\langle g,m|.
\end{eqnarray}

Three sequential interactions (the first one with quantum field in
the cavity in the first interaction region, the second with
intermediate classical field transforming $|g\rangle$ into the
auxiliary state $|f\rangle$, which is off-resonant, and finally
the third with quantum field) will transform to
\begin{eqnarray}
|g,n\rangle &&\longrightarrow
|f,n\rangle\cos(\phi\sqrt{n})\\
&&-\frac{1}{2}i|e,n-1\rangle \sin(2\phi\sqrt{n}) - |g,n\rangle
\sin^2(\phi\sqrt{n}).\nonumber
\end{eqnarray}

This transformation is considered ideal here, though finite
efficiency can also easily be accounted for. If as in micromaser
theory we are interested only in the evolution of the state in the
cavity, ignoring the states of the atoms, we can trace over the
atom states $\rho(n)\to tr_{atom}{P_{after}}$, obtaining after
some transformations
\begin{eqnarray}
\rho &&\longrightarrow \cos^2(\phi
\sqrt{n})\rho(n)\\
&&+\sin^4(\phi \sqrt{n})\rho(n)+ \frac{1}{4}\sin^2(2\phi \sqrt{n
+1})\rho(n+1).\nonumber
\end{eqnarray}

The first term here states that the atom left the first
interaction unexcited and was transformed to the auxiliary state,
the second one that atom absorbs the photon and returns it back to
the cavity, and the third is the probability of the atom leaving
the cavity in excited state.

Taking into account the interaction of the cavity field with the
heat-bath between atom flights \cite{Englert, Meystre}, we obtain
the following master equation
\begin{eqnarray}
\frac{\partial \rho}{\partial t} =&&-\frac{r}{4}\sin^2(2\phi
\sqrt{n})\rho(n) +\gamma\bar n n\rho(n-1)\\
&&-\gamma(\bar n+1)n \rho(n) +\frac{r}{4}\sin^2(\phi \sqrt{n+1})\rho(n+1)
\nonumber\\&&-\gamma\bar n(n+1) \rho(n)+\gamma(\bar n+1)(n+1)\rho(n+1).\nonumber
\end{eqnarray}
Here $\gamma=\frac{\omega}{2\pi Q}$ is the decay constant of the
cavity, and $\bar n=[\exp(\hbar\omega/kT)-1]^{-1}$ is the mean
number of photons in the cavity for the thermal state at
temperature $T$. Each of the 6 terms in the master equation
corresponds to the probability of transition to or from the level
$n$ caused by the atom or heat-bath. The first three terms can be
transformed to the last three terms with the replacement $n\to
n+1$ and a change of sign. Taking into account that these first
three terms also become zero for $n=0$, one can easily obtain a
steady-state solution for the QBD when $\frac{\partial
\rho}{\partial t}=0$:
\begin{eqnarray}
\rho_{SS}(n)&&= \\
&&\rho_{SS}(0)\prod\limits^{n}_{m=1}\frac{4\gamma \bar n m}{r
\sin^2(2\phi \sqrt{m}) + 4\gamma (\bar n+1)m}. \nonumber
\label{antimaser}
\end{eqnarray}

The results of numerical calculations of the mean photon number
$\langle n\rangle$ and relative variance (Fano factor)
$Q_f=(\langle n^2\rangle-\langle n\rangle^2)/\langle n\rangle$
according to (\ref{antimaser}) for the  set of parameters
achievable in the experiment are presented in Fig. 2. As is
clearly seen, the QBD works as an effective cooler for the cavity
field. For most of the Rabi phase values the state in the cavity
is close to the vacuum state, except peaks corresponding to
multiples of $\pi/2$. For these values of the phase sub-Poissonian
statistic of the field ($Q_f<1$) is observed with minima of the
Fano factor. This is the regime of the QND state selector. Large
maxima of the Fano factor correspond to Rabi phase resonance for
two-photon states and small maxima to three-photon states
(correspondingly, $\phi = k \pi/(2 \sqrt{2})$ and $\phi = k \pi/(2
\sqrt{3}$).

\begin{figure}
\includegraphics[width=0.47 \textwidth]{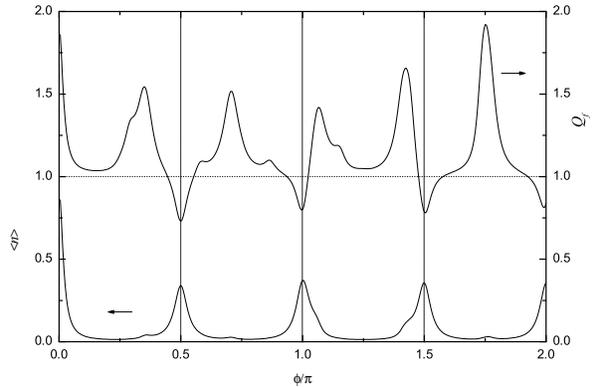}
\caption {\small Dependence of steady states of e.m. field in the
QBD on the Rabi phase $\phi$ with atom rate $r=3000$ s$^{-1}$,
temperature $T=1.4$ K ($\bar n=0.92$) and $Q=2\times10^9$. The
lower curve is the mean number of photons $\langle n\rangle$, the
upper curve is the Fano factor $Q_f$.} \label{fig2}
\end{figure}

If we now shall start monitoring the state of the atoms leaving
the cavity with the velocity chosen to satisfy condition $\phi =
\pi/2$, we can distinguish in a quantum-nondemolition way two
lowest quantum $n$-states $|0\rangle$ and $|1\rangle$ - quantum
bit. If dissipation in the system is absent, every new measurement
of the state (every new atom) will provide the same result as the
first measurement since QND measurements are repetitive and the
initial state will be preserved. If the Q-factor is limited and
the temperature is not zero, then due to interaction with the
heat-bath the cavity can lose and acquire photons and the lifetime
of $n$ states is limited; if however the rate of atoms
$r\gg\gamma$, repetitive measurements are still possible and,
plotting the dependence of the atom state on the time of
measurement, we'll observe characteristic steps of comparable
length, corresponding to thermal transitions between the
$|0\rangle$ and $|1\rangle$ states.

Fig. 3 presents the results of a Monte-Carlo simulation of this
possible experiment.

\begin{figure}
\includegraphics[width=0.47 \textwidth]{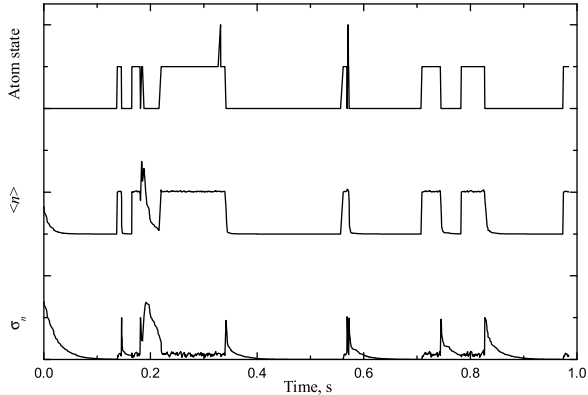}
\caption {\small Simulation of quantum bit detection regime at
$\phi=\pi/2$. The parameters used in the calculations are the same
as in Fig. 2. The upper curve is the state of the atoms detected,
the central curve is the mean number of photons in the cavity
$\langle n\rangle$, the lower curve is the standard deviation of
the QBD $\sigma_n=\sqrt{\langle n^2\rangle - \langle
n\rangle^2}$.} \label{fig3}
\end{figure}

Toroidal cavities with the $TE(001)$ mode were found to be most
suitable for the experiment. The lines of electric field are
wrapped around the center, vanishing on the cavity walls. The mode
provides good atom--field coupling (Rabi frequency 47kHz) and has
axial symmetry avoiding difficulties with mode orientation. The
geometrical factor of the cavity is $\Gamma=408\,\Omega$. Several
cavities made of pure niobium (99.9\%) were manufactured and
tested. The cavities comprised two parts, a cup and a cover,
welded by electron beam welding, chemically etched and baked in
ultra high vacuum at 1800$^\circ$C. The cover has a membrane for
tuning the resonant frequency by mechanical and piezo squeezing.
The internal dimensions of the cavities are as follows: inner
diameter  -- 21 mm;  outer diameter -- 38 mm; height -- 12.7 mm.
The cavities were tested at temperature 1.4\,K and a Q-factor
$2\times10^9$ was observed. These obtained parameters were used in
the theoretical calculations and simulations of the previous
section. The measured frequency intervals for mechanical
adjustment of the cavities (10 MHz) and piezo fine tuning (250
KHz) are appropriate for the achieved precision of cavity
manufacturing.

The experimental setup for the QBD in general is similar to that
described in Refs. \cite{Meschede, UFN} and consists of a pumped
$4He$ cryostat (achievable $T=1.3$ K) and laser system. A beam of
$Rb$ atoms is produced in an atomic oven connected to the
cryostat. The cavity and tuning mechanism are fixed to the
coldfinger attached directly to the helium bath of the cryostat.
The state-selective atom detector, mounted a few centimeters
behind the resonator, allows atoms to be detected atoms either in
the ground and excited states or in the auxiliary and ground
states. It consists of an electrostatic system creating gradient
ionizing field and two channeltron electron detectors
\cite{Meschede, Varcoe}.

The laser system for preparing the Rydberg excited state is based
on a cw ring dye laser and external stabilized intracavity
frequency doubler with UV power output about 15\,mW at wavelength
of 297nm of . Since the experiment requires a defined interaction
time, Doppler velocity selective excitation is employed with the
laser beam inclined at an angle of about $11^\circ$ to the normal
angle. The laser frequency is stabilized on the same Rydberg
transition using an auxiliary chamber with atomic beam.

To prepare atoms in the ground state, required for the final QBD
experiment, the same laser setup can be used with the addition of
an auxiliary Stark field to allow forbidden $5^2S_{1/2}\to 61^2
D_{5/2}$ transitions.

In preliminary experiments the beam of Rydberg atoms in the
excited state was guided through the cavity and detected. Count
rates up to 30,000 atoms/sec were measured.

To sum up, the parameters already achieved in the experiment open
the possibility of demonstrating repetitive QND detection of
vacuum and single-photon states of a microwave field.

The authors are grateful to Professor V.B.Braginsky for his
supporting interest in this experiment. The work of M.L.Gorodetsky
was supported by an Alexander von Humboldt research fellowship.

\end{document}